# THz-PINNs: Time-Domain Forward Modeling of Terahertz Spectroscopy with Physics-Informed Neural Networks

Pengfei Zhu, *Graduate Student Member, IEEE*, and Xavier Maldague, *Life Senior Member, IEEE*

*Abstract*—**Terahertz time-domain spectroscopy (THz-TDS) is a powerful tool for extracting optical and electrical parameters, as well as probing internal structures and surface morphology of materials. However, conventional simulation techniques, such as the finite element method (FEM) and finite-difference time-domain (FDTD), face challenges in accurately modeling THz wave propagation. These difficulties arise from the broad operating bandwidth of THz-TDS (~0.1-10 THz), which demands extremely high temporal resolution to capture pulse dynamics and fine spatial resolution to resolve microstructures and interface effects. To address these limitations, we introduce physics-informed neural networks (PINNs) into THz-TDS modeling for the first time. Through a comprehensive analysis of forward problems in the time domain, we demonstrate the feasibility and potential of PINNs as a powerful framework for advancing THz wave simulation and analysis.**

## I. INTRODUCTION

Terahertz (THz) radiation is electromagnetic in nature and is in the frequency range of 0.1-10 THz. This lies between the infrared and microwave regions of the electromagnetic spectrum in an area that has been labelled the 'terahertz gap' due to a lack of sufficiently powerful sources that worked at room temperature [1], [2], [3]. The simulation of THz propagation within matter can help us understand the underlying interaction mechanisms (reflection, absorption, and dispersion), especially in biomedical [4], non-destructive testing (NDT) [5], communications fields [6].

A dispersive medium is described by frequency-dependent material parameters: the absorption coefficient, $\alpha(\omega)$, and index of refraction $n(\omega)$ [7]. These parameters can be obtained through EM pulse propagation experiments using Fourier transform (FT). However, FT assumes that the material is a linear, time-invariant (steady-state), and spatially uniform system over the measurement time [8]. In contrast, it is not

This work was supported in part by the Natural Sciences and Engineering Research Council of Canada (NSERC) through the CREATE-oN DuTy! Program under Grant 496439-2017, in part by the Canada Research Chair in Multi-polar Infrared Vision (MIVIM). (*Corresponding author: Pengfei Zhu.*)

Pengfei Zhu and Xavier Maldague are with the Department of Electrical and Computer Engineering, Computer Vision and Systems Laboratory (CVSL), Laval University, Québec G1V 0A6, Québec city, Canada (e-mail: pengfei.zhu.1@ulaval.ca; xavier.maldague@gel.ulaval.ca).

suitable for pump-probe and transient THz spectroscopy experiments [9]. Because the material's optical properties can change rapidly during the THz pulse duration, i.e., photoexcited carrier densities can evolve on ultrafast timescales, altering permittivity dynamically. In addition, the photoexcited region in the sample often has an uneven spatial distribution. This leads to position-dependent variations in optical properties encountered by the THz pulse. These cases are often used to study the dynamic behavior and properties of materials under unstable state, such as understanding ultrafast dynamic processes in materials [10], exploring the optoelectronic properties and device potential of materials [11], and studying the light-matter interactions [12]. The finite-difference time-domain (FDTD) methods are commonplace for solving complex problems for electromagnetic (EM) pulse propagation [13], [14]. The applications of FDTD methods for simulating magnetic and electric nonlinearities of other materials have been discussed [15], [16], [17], [18], [19]. The finite-element method (FEM) is also a useful and powerful computational technique for modelling how electromagnetic waves propagate and interact with their surroundings [20]. Different from FDTD, FEM can solve the EM propagation problem in both time-domain and frequency-domain [21], [22]. As FEM modelling has been applied to study EM propagation in both the microwave and optical portions of the spectrum [23], [24], [25], it is a logical step to use this technique for the study of wave propagation at terahertz frequencies. However, to accurately simulate EM wave propagation, both FDTD and FEM face a stringent spatial resolution constraint: the largest element size must be no larger than one-third of the radiations' wavelength. In many high-fidelity models, this is further reduced to one-tenth of the radiation wavelength to ensure numerical accuracy [26]. For instance, at 1 THz (wavelength ~300 µm), this results in a critical element size of ~30 µm. When applied to a non-destructive testing (NDT) sample with dimensions of $100\times100\times5$ mm³, the simulation requires an extremely large number of mesh elements – resulting in significantly computational costs in terms of memory and processing time. Moreover, FDTD has several inherent limitations: it typically requires uniform grids, which are inefficient for modelling fine features or localized structures; it struggles with highly dispersive or anisotropic materials unless advanced extensions are introduced; and its time-domain nature makes it less efficient for narrowband or steady-state problems where frequency-domain solvers like FEM are preferable. Additionally, FDTD simulations become unstable/accurate when dealing with materials whose EM properties change rapidly in time/space.

Physics-informed learning [27], introduced in a series of papers by Karniadakis's group both for Gaussian-process regression [28] and physics-informed neural networks (PINNs) [29], can seamlessly integrate multifidelity / multimodality experimental data with the various partial differential equations (PDEs) for solid/fluid mechanics [30], [31], quantum mechanics [32], heat transfer [33], etc. PINNs offer a promising alternative to traditional numerical solvers like FDTD and FEM for modelling EM wave propagation in the THz regime. Unlike mesh-based methods, PINNs solve PDEs by incorporating physical laws directly into the loss function of a neural network, enabling solutions without explicit meshing. This mesh-free nature allows PINNs to handle complex geometries, heterogeneous media, and high-dimensional problems more flexibly and efficiently.

In this work, we address the forward problem of THz wave propagation by applying physics-informed neural networks (PINNs) to the terahertz regime for the first time. Specifically, two time-domain Maxwell's equations, covering both conductive and lossless media, are embedded into the neural network framework, leading to the development of THz physics-informed neural networks (THz-PINNs). The proposed networks are validated against numerical simulations based on the finite-difference time-domain (FDTD) method. Material properties are further characterized through THz-TDS experiments. Additionally, we incorporate the auxiliary differential equation (ADE) method with the Drude-Lorentz model as novel loss functions and feed into the networks for training. This study represents the first attempt to address THz-TDS modeling using a mesh-free, physics informed framework.

## II. THEORY

The principle of terahertz time-domain spectroscopy follows the Maxwell equation:

$$\nabla \cdot E = \frac{\rho}{\varepsilon} \tag{1}$$

$$\nabla \cdot B = 0 \tag{2}$$

$$\nabla \times E = -\frac{\partial B}{\partial t} \tag{3}$$

$$\nabla \times B = \mu J + \mu \varepsilon \frac{\partial E}{\partial t} \tag{4}$$

where $E$ denotes the electrical field vector, $B$ denotes the magnetic flux density, $\rho$ denotes the free charge density, $J$ denotes the free current density vector, $\varepsilon$ denotes the permittivity of the medium, $\mu$ denotes the permeability of the medium, and $t$ denotes the time. The above equations can be arranged as:

$$\nabla^2 E - \mu \varepsilon \frac{\partial^2 E}{\partial t^2} = \mu \frac{\partial J}{\partial t} + \nabla(\frac{\rho}{\varepsilon}) \tag{5}$$

Assuming there is no free charge and current, i.e., $\rho = 0$ and $J = 0$, the equation is degraded as:

$$\nabla^2 E = \mu \varepsilon \frac{\partial^2 E}{\partial t^2} = \frac{1}{c^2} \frac{\partial^2 E}{\partial t^2} \tag{6}$$

where $c$ is the light speed. In a source-free and lossless medium, $c = 1/\sqrt{\mu \varepsilon}$. However, most of materials are highly dispersive in the terahertz band. Therefore, in the time-domain, there is a damping term proportional to $\partial E/\partial t$:

$$\nabla^2 E - \mu \varepsilon \frac{\partial^2 E}{\partial t^2} = \mu \sigma \frac{\partial E}{\partial t} \tag{7}$$

where $\sigma$ is the conductivity ($J = \sigma E$). There are three parameters in Eq. (7).

## III. TERAHERTZ PHYSICS-INFORMED NEURAL NETWORKS

The physics-informed neural networks (PINNs) are employed for the first time, as shown in Fig. 1. The loss function serves as a quantitative measure of the discrepancy between the network predictions and the reference data. The objective of the training process is to minimize the loss, thereby enhancing the predictive accuracy of the networks. For PINNs, the loss function is formulated as:

$$\mathcal{L} = \mathcal{L}_{PDE} + \mathcal{L}_{IC} + \mathcal{L}_{BC} \tag{8}$$

where $\mathcal{L}_{PDE}$ is the loss function for PDE, $\mathcal{L}_{IC}$ is the loss function for initial condition, and $\mathcal{L}_{BC}$ is the loss function for boundary condition. For a source-free and lossless medium, the $\mathcal{L}_{PDE}$ becomes:

$$\mathcal{L}_{PDE} = \frac{1}{N_f} \sum_{i=1}^{N_f} (\nabla^2 E - \frac{n^2}{c^2} \frac{\partial^2 E}{\partial t^2})^2 \tag{9}$$

where $n$ is the refractive index of the medium. Because there is no initial electric field in THz-TDS, the $\mathcal{L}_{IC}$ becomes:

$$\mathcal{L}_{IC} = \frac{1}{N_{IC}} \sum_{i=1}^{N_{IC}} (\left| E(x^i, y^i, 0) \right|^2 + \left| \frac{\partial E(x^i, y^i, 0)}{\partial t} \right|^2) \tag{10}$$

In conventional FDTD and FEM approaches, boundary conditions must be specified over a predefined computational domain, which may encompass the entire sample. In this work, to simplify the formulation, the boundary condition is defined as:

$$\mathcal{L}_{BC} = \mathcal{L}_{Left} + \mathcal{L}_{Right} + \mathcal{L}_{Up} + \mathcal{L}_{Down} \tag{11}$$

$$\mathcal{L}_{Left} = \frac{1}{N_{Left}} \sum_{i=1}^{N_{Left}} |E(0, y^i, t^i) - f(y^i, t^i)|^2 \tag{12}$$

$$\mathcal{L}_{Right} = \frac{1}{N_{Right}} \sum_{i=1}^{N_{Right}} |\frac{\partial E(L_x, y^i, t^i)}{\partial t} + \frac{c}{n} \frac{\partial E(L_x, y^i, t^i)}{\partial x}|^2 \tag{13}$$

$$\mathcal{L}_{Up} = \frac{1}{N_{Up}} \sum_{i=1}^{N_{Up}} |\frac{\partial E(x^i, L_y, t^i)}{\partial t} + \frac{c}{n} \frac{\partial E(x^i, L_y, t^i)}{\partial y}|^2 \tag{14}$$

$$\mathcal{L}_{Down} = \frac{1}{N_{Down}} \sum_{i=1}^{N_{Down}} |\frac{\partial E(x^i, 0, t^i)}{\partial t} - \frac{c}{n} \frac{\partial E(x^i, 0, t^i)}{\partial y}|^2 \tag{15}$$

where $f(y^i, t^i)$ is the input electromagnetics wave (optical source). In conventional simulation methods, we often set pre-defined (e.g., Gaussian-shaped) input for simulation, which is different from real input. $\mathcal{L}_{BC}$, $\mathcal{L}_{IC}$, $\mathcal{L}_{PDE}$ penalize the residuals, that is, the difference between theoretically correct values and network predicted values of the boundary conditions, initial conditions, and governing equations, respectively. $N_{IC}$, $N_{Left}$, $N_{Right}$, $N_{Up}$, and $N_{Down}$ are the numbers of data points for different terms.

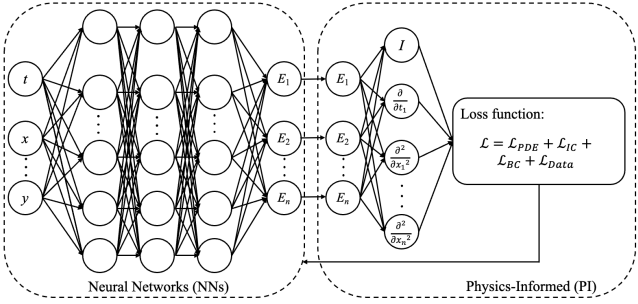

**Fig. 1.** Schematic image of THz-PINNs.

A key advantage of simulation-assisted PINNs is the ability to incorporate pre-known data into the network, which accelerates convergence and improves training efficiency. Accordingly, an additional data-driven loss term is introduced, constraining the network output to match the response governed by the Maxwell equations under investigation:

$$\mathcal{L}_{Data} = \frac{1}{N_{Data}} \sum_{i=1}^{N_{Data}} |E(x^i, y^i, t^i) - E_{Data}(x^i, y^i, t^i)|^2 \quad (16)$$

where $N_{IC}$ is the number of data points for data loss $\mathcal{L}_{Data}$. This type of computation is achieved in the PINNs framework using automatic differentiation [34], an operation which is a key enabler for PINNs. It combines the derivatives of the constituent operations using the chain rule and outputs the derivative of the overall composition, defined as the entire sequence of mathematical operations that constitute the neural network's forward propagation.

For a medium with the conductivity, the $\mathcal{L}_{PDE}$ becomes:

$$\mathcal{L}_{PDE} = \frac{1}{N_f} \sum_{i=1}^{N_f} (\nabla^2 E - \mu\varepsilon \frac{\partial^2 E}{\partial t^2} - \mu\sigma \frac{\partial E}{\partial t})^2 \quad (17)$$

The initial condition is identical to Eq. (10). The boundary conditions become:

$$\mathcal{L}_{BC} = \mathcal{L}_{Left} + \mathcal{L}_{Right} + \mathcal{L}_{Up} + \mathcal{L}_{Down} \quad (18)$$

$$\mathcal{L}_{Left} = \frac{1}{N_{Left}} \sum_{i=1}^{N_{Left}} |E(0, y^i, t^i) - f(y^i, t^i)|^2 \quad (19)$$

$$\mathcal{L}_{Right} = \frac{1}{N_{Right}} \sum_{i=1}^{N_{Right}} |\frac{\partial E(L_x, y^i, t^i)}{\partial t} + \frac{1}{\sqrt{\mu\varepsilon}} \frac{\partial E(L_x, y^i, t^i)}{\partial x}|^2 \quad (20)$$

$$\mathcal{L}_{Up} = \frac{1}{N_{Up}} \sum_{i=1}^{N_{Up}} |\frac{\partial E(x^i, L_y, t^i)}{\partial t} + \frac{1}{\sqrt{\mu\varepsilon}} \frac{\partial E(x^i, L_y, t^i)}{\partial y}|^2 \quad (21)$$

$$\mathcal{L}_{Down} = \frac{1}{N_{Down}} \sum_{i=1}^{N_{Down}} |\frac{\partial E(x^i, 0, t^i)}{\partial t} - \frac{1}{\sqrt{\mu\varepsilon}} \frac{\partial E(x^i, 0, t^i)}{\partial y}|^2 \quad (22)$$

It is possible to find that the material parameters are $\varepsilon$ (permittivity) and $\mu$ (permeability) rather than the light speed. All models were implemented with the PyTorch framework, a code database in Python software, specifically developed for achieving deep learning, and then trained using NVIDIA 4060 Titan GPUs. After a lot of attempts, we select the Tanh as the activation function. The network layers are 8, including the input layer, 6 hidden layers, and 1 output layer. Where each hidden layer contains 128 neurons. To balance the contributions of different loss terms in training THz-PINNs, we adopt a gradient-based adaptive weighting scheme. For each loss term $\mathcal{L}_k$, we compute the gradient norm $g_k = ||\nabla_\theta \mathcal{L}_k||$ and assign the weight $\lambda_k = \frac{1/g_k}{\sum_j 1/g_j}$. This ensures that loss terms with small gradients are amplified, preventing them from being neglected during training.

## IV. EXPERIMENTS AND SIMULATIONS

### A. Experimental setup

To validate the computational accuracy of PINNs, the simulation results are compared against both PINNs predictions and experimental measurements. It is noted that THz-TDS can only capture the external electromagnetics wave; therefore, external data are incorporated into the neural networks for training.

The THz-TDS imaging system is shown in Fig. 2. An ultra-fast laser pulse is split into a pump beam and a reference beam. The pump beam is time-delayed via an optical delay line and directed to a THz emitter, which generates linearly polarized THz radiation. This THz wave passes through the sample and is collected by a detector. The reference beam served as the sampling signal at the detector. The sampled signal is then processed by a lock-in amplifier to enhance weak signals for data acquisition. The THz system was manufactured by Menlo Systems GmbH, Munich, Germany. It features a frequency resolution of 1.2 GHz and a repetition rate of 100 MHz. The experiments were conducted in transmission mode with a scanning step of 0.5 mm. Additionally, the ambient temperature was controlled at 22 °C ± 0.1 °C, with relative humidity maintained at 50% ± 2 %.

Generally, optical parameter measurements are performed in transmission to minimize the influence of alignment errors. The refractive index $n(\omega)$ can then be extracted from the phase difference between the sample and reference signals [35], [36]:

$$n(\omega) = 1 + \frac{c}{2\pi\omega d} (\varphi_{sam}(\omega) - \varphi_{ref}(\omega)) \quad (23)$$

where $\varphi_{sam}(\omega)$ and $\varphi_{ref}(\omega)$ are the phase angles of the sample signal and reference signal, $c$ is the speed of the light, $\omega$ is the frequency and $d$ is the sample thickness. The absorption coefficient can be calculated as:

$$\alpha(\omega) = -\frac{2}{d} \ln \left[ r(\omega) \frac{(n(\omega)+1)^2}{4n(\omega)} \right] \quad (24)$$

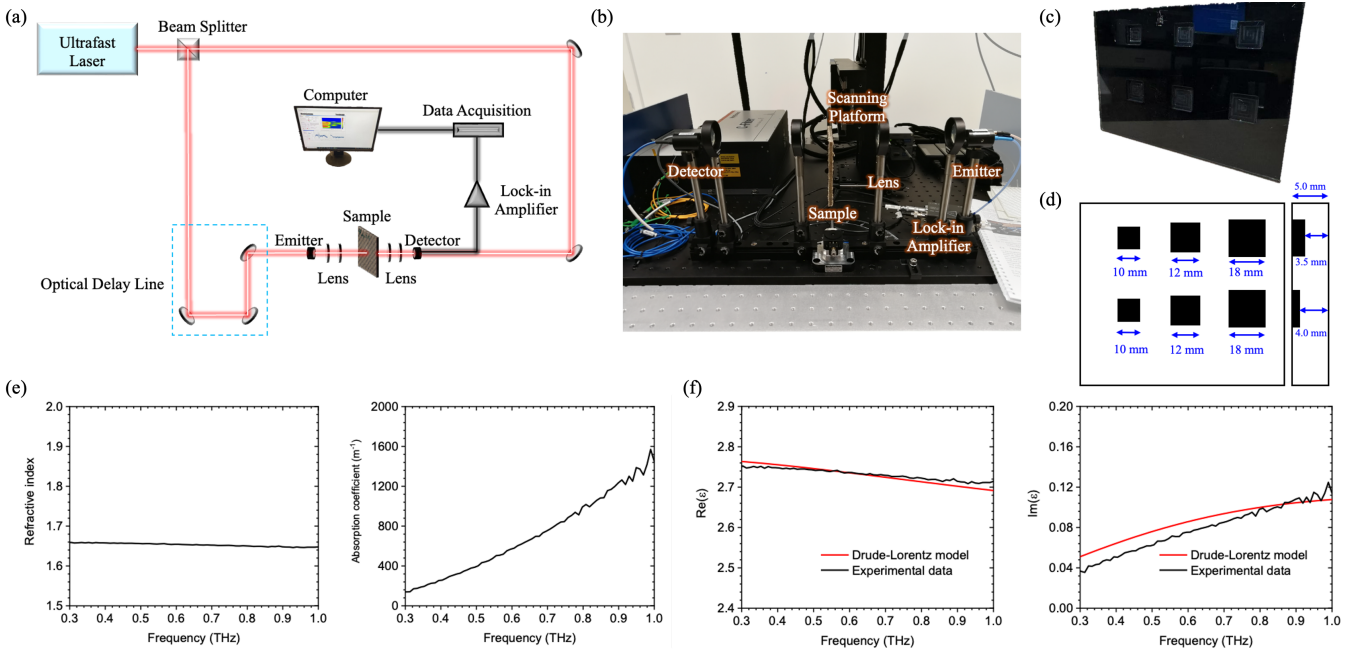

**Fig. 2.** Experimental systems and the sample: (a) Schematic image of experimental systems; (b) Photograph of experimental systems; (c) Photograph of the sample; (d) The schematic image of the sample; (e) The refractive index and absorption coefficient of the sample; (f) The Drude-Lorentz model and experimental data for the real (Re) and imaginary (Im) parts of permittivity.

where $r(\omega) = E_{sam}(\omega) / E_{ref}(\omega)$ is the ratio between sample signal $E_{sam}(\omega)$ and reference signal $E_{ref}(\omega)$. In addition to optical parameters, the dielectric constant is also used to characterize the tested sample. To consider the material dispersion, a Drude-Lorentz model can be employed with multiple coefficient:

$$\varepsilon(\omega) = \varepsilon_\infty - \frac{\omega_p^2}{\omega^2 + i\gamma\omega} + \sum_{j=1}^{N} \frac{\Delta\varepsilon_j \omega_{0,j}^2}{\omega_{0,j}^2 - \omega^2 - i\Gamma_j\omega} \quad (25)$$

where $\varepsilon_\infty$ is the high-frequency dielectric constant, $\omega_p$ is the plasma frequency, $\gamma$ is the damping factor, $\omega_{0,j}$ is the resonance frequency of the $j$-th oscillator, $\Delta\varepsilon_j$ is the strength of the $j$-th resonance, $\Gamma_j$ is the damping coefficient of the $j$-th resonance.

### B. Simulation: Finite-Difference Time-Domain

In a 2D Cartesian coordinate system, substituting central-difference approximations for the spatial and temporal derivatives into Maxwell's equations yields coupled finite-difference time-stepping relations for the electric field $E$ and magnetic field $H$ components [37], [38], [39], [40]. The time-stepping scheme for the field components in a 2D model can be written as:

$$\left(1 + \frac{\Delta t \cdot \sigma}{2\varepsilon}\right)(E_x)_{(i+\frac{1}{2},j)}^{n+1} = \left(1 - \frac{\Delta t \cdot \sigma}{2\varepsilon}\right)(E_x)_{(i+\frac{1}{2},j)}^{n} + \frac{\Delta t}{\delta\varepsilon}[(H_z)_{(i+\frac{1}{2},j+\frac{1}{2})}^{n+\frac{1}{2}} - (H_z)_{(i+\frac{1}{2},j-\frac{1}{2})}^{n+\frac{1}{2}}] \quad (26)$$

$$\left(1 + \frac{\Delta t \cdot \sigma}{2\varepsilon}\right)(E_y)_{(i,j+\frac{1}{2})}^{n+1} = \left(1 - \frac{\Delta t \cdot \sigma}{2\varepsilon}\right)(E_y)_{(i,j+\frac{1}{2})}^{n} - \frac{\Delta t}{\delta\varepsilon}[(H_z)_{(i+\frac{1}{2},j+\frac{1}{2})}^{n+\frac{1}{2}} - (H_z)_{(i-\frac{1}{2},j+\frac{1}{2})}^{n+\frac{1}{2}}] \quad (27)$$

$$\left(1 + \frac{\Delta t \cdot \sigma}{2\mu}\right)(H_x)_{(i,j+\frac{1}{2})}^{n+1} = \left(1 - \frac{\Delta t \cdot \sigma}{2\mu}\right)(H_x)_{(i,j+\frac{1}{2})}^{n-\frac{1}{2}} + \frac{\Delta t}{\delta\mu}[(E_y)_{(i,j+\frac{1}{2})}^{n} - (E_y)_{(i,j)}^{n}] \quad (28)$$

$$\left(1 + \frac{\Delta t \cdot \sigma}{2\mu}\right)(H_y)_{(i+\frac{1}{2},j)}^{n+\frac{1}{2}} = \left(1 - \frac{\Delta t \cdot \sigma}{2\mu}\right)(H_y)_{(i+\frac{1}{2},j)}^{n-\frac{1}{2}} - \frac{\Delta t}{\delta\mu}[(E_x)_{(i+\frac{1}{2},j)}^{n} - (E_x)_{(i,j)}^{n}] \quad (29)$$

where $i$, $j$, and $k$ represent the nodes of the Yee cells, and $n$ represents the calculated time step. Equations (26)-(27) represent the electric field components, which are located on the three edges at each Yee cell. Equations (28)-(29) represent the magnetic field components, which are located in the middle of the three surfaces. In addition to the one-half spatial-cell displacement between $E$ and $H$, there is also one-half time-cell displacement.

### C. Sample

The sample under investigation is a glass fiber-reinforced polymer (GFRP) laminate containing six prefabricated defects, as shown in Fig. 2(c). The top three-square defects have a depth of 3.5 mm, while the bottom three have a depth of 4.0 mm. From left to right, the side lengths of defects are 10, 12, and 18 mm, respectively. A schematic of the defect distribution is shown in Fig. 2(d). The refractive index and absorption coefficient are calculated using Eq. (23) and (24), with the results shown in Fig. 2(e). Furthermore, the Drude-Lorentz model in Eq. (25) is used to fit the experimental data, yielding the real and imaginary parts of the permittivity, as shown in Fig. 2(f).

## V. Results and Discussion

### A. Non-Dispersive Medium

For lossless medium, the only material parameter in Eq. (6) and Eq. (9) is the refractive index. The refractive index obtained from the transmission-mode THz-TDS experiments (see Fig. 2(e)) is used as input for the FDTD simulations. In contrast, the initial refractive index in the THz-PINNs is set to 1. The total training epochs were set to 50000. Due to the large amount of data generated by meshing, the array was subsampled with a step size of 5, and 20% of the randomly selected data points were used to accelerate the training. In addition, both spatial and temporal variables were normalized to values between 0 and 1, and the speed of light was normalized to 1.

The prediction results obtained from THz-PINNs are shown in Fig. 3. Compared with the FDTD results, the THz-PINNs exhibit strong consistency, accurately describing the propagation of the THz pulse from right to left. However, the THz-PINNs fail to reproduce the secondary scattering waves and boundary absorption at the sample's boundary. This limitation arises because only simple absorbing boundary conditions, as defined in Eqs. (13)-(15), were imposed in the THz-PINNs, and the incident wave was introduced from the right boundary rather than from outside the computational domain. The maximum error between FDTD and THz-PINNs is no more than 0.25 except for 55 ps.

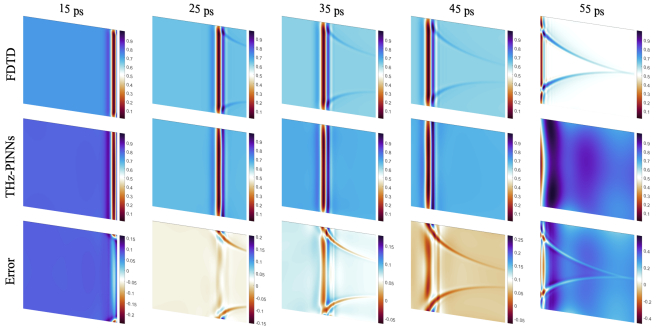

**Fig. 3.** Comparing the prediction results of THz-PINNs with FDTD in non-dispersive medium.

The computational time is also a significant index to evaluate conventional FDTD solvers and THz-PINNs. Here, the computational time for 2D FDTD solvers is just 60 s. As for THz-PINNs, due to introducing 20% simulation data, the training process was significantly accelerated. The computational time of THz-PINNs in 2D non-dispersive medium is just 5 minutes. Of note, this is significantly different from conventional simulation solvers. PINNs require only once training, then they can predict multiple parameters / data points in the same model without re-computing.

### B. Dispersive Medium

In general THz-TDS systems, the THz source is broadband; thus, the narrowband approximation is not suitable, as it neglects the frequency-dependent behavior of optical parameters. Existing approaches can be broadly classified into two categories: 1) transforming the time-domain Maxwell equations into frequency-domain Helmholtz equations; and 2) using the auxiliary differential equation (ADE) method. Both techniques introduce additional equations and variables, thereby increasing the computational cost and reducing convergence accuracy.

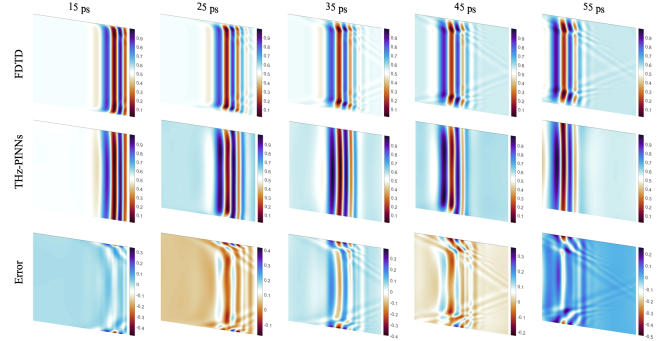

**Fig. 4.** Comparing the prediction results of THz-PINNs with FDTD in dispersive medium.

Similarly, the THz-PINNs were employed to simulate the THz wave propagation in dispersive medium. The total training epochs were set to 50000. The loss terms were changed to the dispersive medium with a damping term, i.e., using the loss functions in Eqs. (16)-(22). The gradient-adaptive adaptive weighting scheme was used to balance different loss terms. Other training parameters are the same to the previous section.

The simulation results of FDTD and THz-PINNs are shown in Fig. 4. It is clear that THz-PINNs can effectively reconstruct the THz wave propagation in dispersive medium. Comparing with results in Fig. 3, there is strong dispersive phenomenon in Fig. 4. For instance, the THz pulse gradually broadens in time. Although THz-PINNs can accurately predict the movement path of THz waves, the error between THz-PINNs and FDTD cannot be neglected. The maximum errors of each image are more than 0.3. Therefore, an ADE-based THz-PINNs simulation method was proposed for the first time.

THz-PINNs do not require the explicit formulation of all ADEs. Instead, the material parameters in the Drude-Lorentz model ($\varepsilon_\infty$, $\omega_p$, $\gamma$, $\omega_{0,j}$, $\Delta\varepsilon_j$, and $\Gamma_j$) are treated as trainable variables. The neural networks are trained to minimize the residual of Maxwell's equations while adaptively converging to physically consistent values guided by experimental data. Under this framework, the PDE loss is expressed as:

$$\mathcal{L}_{PDE} = \mathcal{L}_{Maxwell} + \mathcal{L}_{P_j(x,t)} \tag{30}$$

$$\mathcal{L}_{Maxwell} = \frac{1}{N_f}\sum_{i=1}^{N_f}(\nabla^2 E - \mu_0\varepsilon_0\varepsilon_\infty\frac{\partial^2 E}{\partial t^2} - \mu_0\sum_j\frac{\partial^2 P_j}{\partial t^2})^2 \tag{31}$$

$$\mathcal{L}_{P_j(x,t)} = \frac{1}{N_f}\sum_{i=1}^{N_f}(\frac{\partial^2 P_j}{\partial t^2} + \gamma_j\frac{\partial P_j}{\partial t} + \omega_{0,j}^2 P_j - \varepsilon_0\Delta\varepsilon_j\omega_{0,j}^2 E)^2 \tag{32}$$

where $P_j$ is the $j$-th polarization, and $\gamma_j$ is the damping coefficient of $j$-th oscillator.

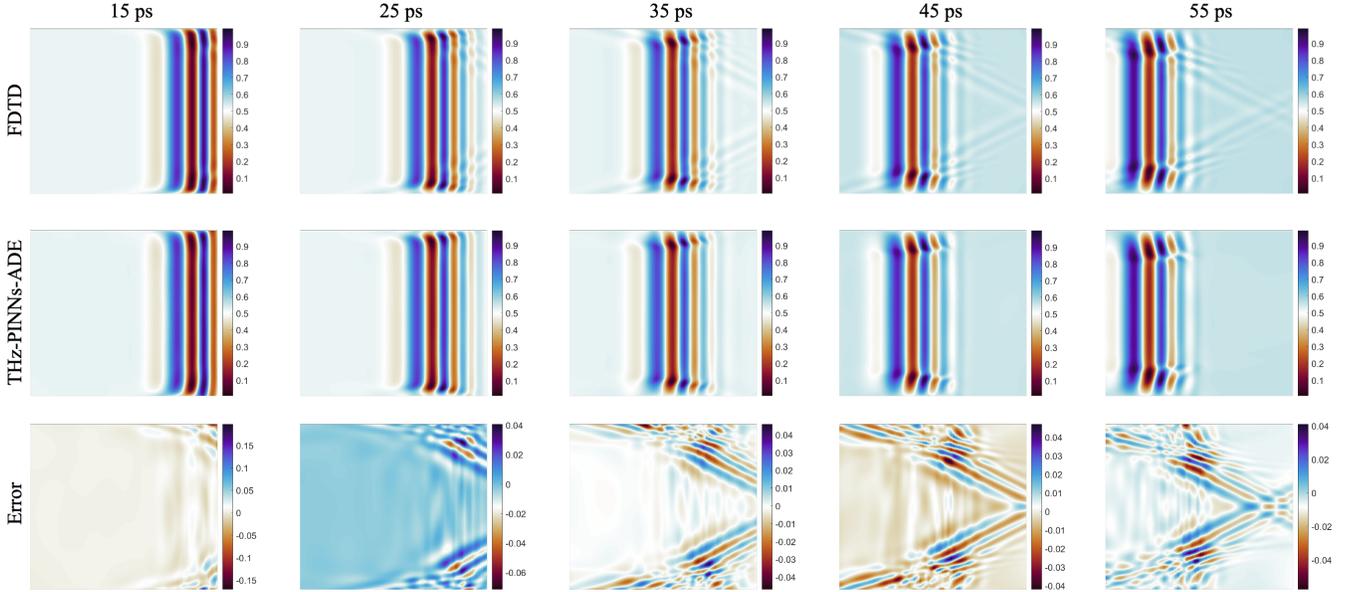

**Fig. 5.** Comparing the prediction results of FDTD (first row) with THz-PINNs-ADE in dispersive medium.

The total training epochs were set to 50000. The hyperparameters, gradient-based adaptive weighting scheme, and data downsampling strategy are identical to those used in the previous training.

The prediction results are shown in Fig. 5. The THz-PINNs-ADE denotes the THz-PINNs combined with the ADE method, i.e., using the PDE loss functions in Eqs. (30)-(32). As shown in Fig. 5, THz-PINNs-ADE can effectively capture the propagation of the incident wave. In addition, THz-PINNs-ADE has higher accuracy than the previous THz-PINNs in Fig. 4. The maximum error extremely decreases and is no more than 0.2.

To quantitatively compare the prediction accuracy between the THz-PINNs and THz-PINNs-ADE, we select the root mean square error as the evaluation index:

$$\text{RMSE} = \sqrt{\frac{1}{MN}\sum_{x=1}^{M}\sum_{y=1}^{N}(I(x,y)-\hat{I}(x,y))^2} \quad (33)$$

where $M$ and $N$ are the total pixel numbers along the length $x$ and width $y$ directions, respectively. $I$ and $\hat{I}$ are simulation results from FDTD and THz-PINNs / THz-PINNs-ADE. The RMSE values at different time are shown in Table 1.

TABLE I
The Evaluation for THz-PINNs and THz-PINNs-ADE Using Root Mean Square Error.

| Time (ps) | THz-PINNs | THz-PINNs-ADE | Improvement |
|---|---|---|---|
| 15 | 2.48% | 0.85% | +1.63% |
| 25 | 7.01% | 0.65% | +6.36% |
| 35 | 6.73% | 0.62% | +6.11% |
| 45 | 4.92% | 0.69% | +4.23% |
| 55 | 5.15% | 0.95% | +4.20% |

It is clear to find that THz-PINNs-ADE has superior performance than THz-PINNs. The RMSE of THz-PINNs-ADE is stable and no more than 1%. Comparing with THz-PINNs, the maximum improvement of THz-PINNs-ADE is 6.36%. However, the disadvantage of THz-PINNs-ADE is the complex parameter setting. For instance, we need to select appropriate number of oscillators. Additionally, the computational time of THz-PINNs-ADE becomes longer. In this study, for three oscillators, the training time of THz-PINNs-ADE is 7 minutes, while the training time of THz-PINNs is 5 minutes. Therefore, in real applications, we should carefully select the number of oscillators and adjust the hyperparameters.

### C. Experimental Validation

THz-TDS systems capture signals only in reflection and transmission modes. Consequently, all sampled signals originate from the sample surfaces, and only boundary data are available for training. Unlike the previous sections, two-dimensional THz-PINNs (depth + time) are employed to simulate THz wave propagation based on a conductive medium model.

The experimental results are presented in Fig. 6. The scanning path includes both air area and sample regions, as shown in Fig. 6(a). The acquired data consist of 288 × 3000 pixels, where 288 corresponds to spatial pixels along the scanning path and 3000 to temporal pixels, as shown in Fig. 6(b). The received THz signals can be divided into a main pulse and echo pulses, with the main pulse representing the first transmitted wave and the echo pulses corresponding to subsequent reflections. The corresponding time-domain spectra are shown in the right panels. Variations in sample thickness induce time delays and amplitude changes in the time-domain signals, as illustrated in Fig. 6(c). For validation of the proposed THz-PINNs, only the duration of the main pulse is considered. The prediction results are presented in Fig. 6(d).

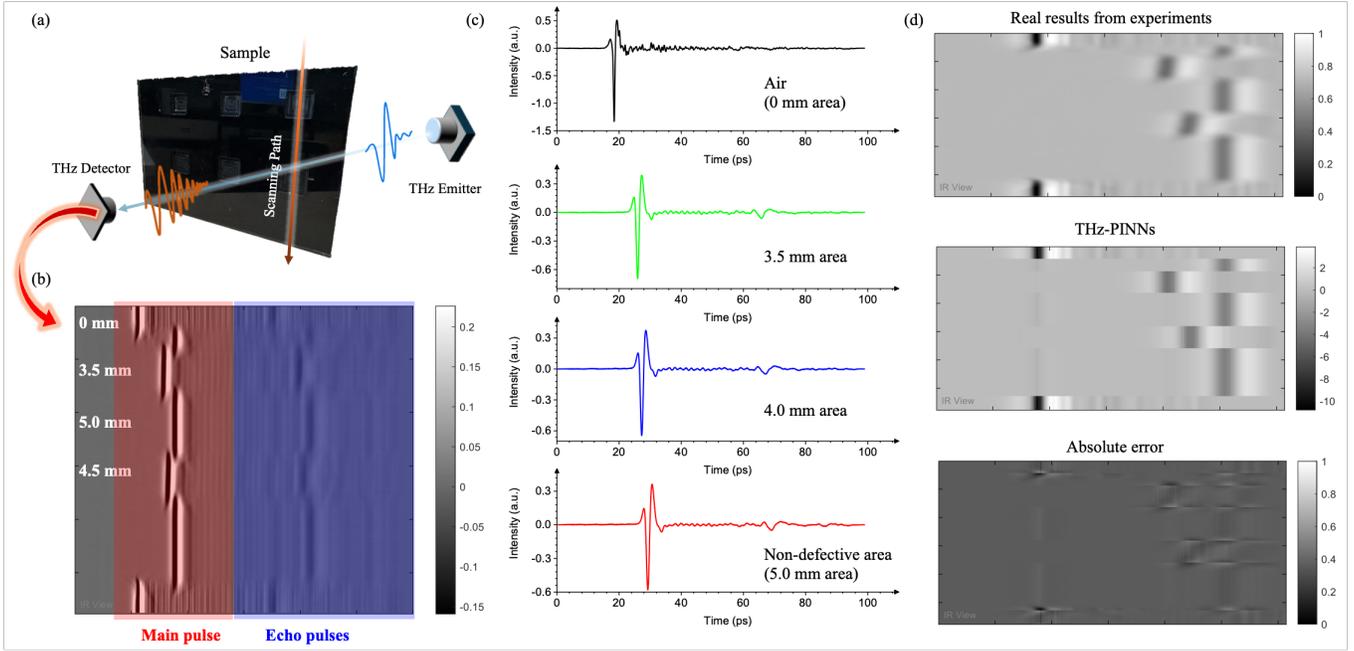

**Fig. 6.** Experimental results: (a) Scanning path of THz-TDS systems; (b) THz signals along scanning direction; (c) THz waveforms of different thickness; (d) Prediction results from THz-PINNs.

THz-PINNs effectively reconstruct the propagation of THz waves across regions of varying thickness. Since the model in this section is two-dimensional, the predictions exhibit sharp boundaries without capturing scattering effects. A quantitative comparison is also presented in Fig. 6(d), indicating that the largest errors occur at the boundary regions.

## VI. CONCLUSION

In this work, we introduce physics-informed neural networks (PINNs) for the first time to simulate terahertz time-domain spectroscopy (THz-TDS), referred to as THz-PINNs. As is well known, THz-TDS simulations typically require a large number of mesh elements, leading to significant computational costs in terms of memory and processing time. THz-PINNs offer a novel approach to simulating the propagation of THz waves, alleviating these challenges. To accelerate the training process, 20% of the simulation data were fed into the network. Comprehensive simulations using both finite-difference time-domain (FDTD) and THz-PINNs were conducted for the forward problem. Specifically, we incorporated Drude-Lorentz model parameters as trainable variables in PINNs and used the auxiliary differential equation (ADE) method to construct new loss functions. Additionally, the feasibility of THz-PINNs was tested with experimental data. The results demonstrate the promising potential of THz-PINNs for simulation and modeling.

However, challenges remain. While THz-PINNs can effectively simulate THz wave propagation, the predicted values near the sample's boundary in THz-PINNs differ from those obtained from FDTD simulations. This is caused by the simple absorbing boundary conditions, as defined in Eqs. (13)-(15). It could be revised / updated in the future work. Furthermore, this work focuses on forward modeling in the time domain. Although the THz-PINNs using Maxwell's equations with a damping term, could predict the movement path of THz waves, the predicted values are different from the results in FDTD. To solve this problem, we introduced the ADE method combined with the Drude-Lorentz model, i.e., THz-PINNs-ADE, which accurately capture the dispersive behavior of THz waves in the medium. The maximum error is no more than 0.2 and the RMSE is no more than 1%. However, we should carefully control the number of oscillators and adjust hyperparameters. In conclude, this work represents the first attempt to PINNs to THz-TDS, validating the feasibility of using PINNs in this domain. Future work will focus on: 1) optimizing the results from THz-PINNs; 2) extending the model to frequency-domain THz wave propagation, 3) solving inverse problems for extraction of material parameters, 4) multi-fidelity PINNs combining simulation and experimental data, 5) potential applications for real-time non-destructive testing.

## ACKNOWLEDGMENT

This work is supported by the Natural Sciences and Engineering Research Council (NSERC) Canada through the Discovery and CREATE 'oN DuTy!' program (496439-2017), as well as the Canada Research Chair in Multipolar Infrared Vision (MiViM).

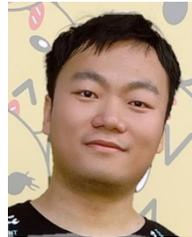

**Pengfei Zhu** received the B.Eng. degree in engineering mechanics from North University of China, Taiyuan, China, in 2019, and the M.Eng. degree in solid mechanics from Ningbo University, Ningbo, China, in 2022. He is currently working toward the Ph.D. degree in electrical engineering with Université Laval, Québec, Canada.

His research interests include non-destructive testing, infrared thermography, deep learning, terahertz time-domain spectroscopy, and photothermal coherence tomography.

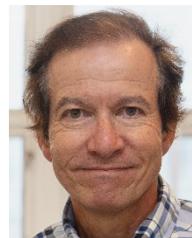

**Xavier Maldague** P.Eng., Ph.D. is full professor at the Department of Electrical and Computing Engineering, Université Laval, Québec City, Canada. He has trained over 50 graduate students (M.Sc. and Ph.D.) and contributed to over 400 publications. His research interests are in infrared thermography, NonDestructive Evaluation (NDE) techniques and vision / digital systems for industrial inspection. He is an honorary fellow of the Indian Society of Nondestructive Testing, fellow of the Canadian Engineering Institute, Canadian Institute for NonDestructive Evaluation, American Society of NonDestructive Testing. In 2019 he was bestowed a Doctor Honoris Causa in Infrared Thermography from University of Antwerp (Belguim).